\documentclass[12pt]{article}
\usepackage{graphicx}
\usepackage{lscape}
\usepackage{amssymb}
\usepackage{pazh}
\tightenlines

\begin{document}

{\footnotesize Astronomy Letters, Vol. 30, No. 12, 2004, pp. 824-833. Translated from Pis'ma v Astronomicheskii 
Zhurnal, Vol. 30, No. 12, 2004, pp. 905­914. Original Russian Text Copyright \copyright\, 2004 by Filippova, 
Lutovinov, Shtykovsky, Revnivtsev, Burenin, Aref'ev, Pavlinsky, Sunyaev.}

\title{\bf Broadband Observations of the Transient X-ray Pulsar \mbox{SAX J2103.5+4545}}

\author{\bf \hspace{-1.3cm} \ \
E.V. Filippova\affilmark{1}, A.A. Lutoviniv\affilmark{1}$^{\,*}$,
P.E. Shtykovsky\affilmark{1}, M.G. Revnivtsev\affilmark{1,2},
R.A. Burenin\affilmark{1}, V.A. Aref'ev\affilmark{1},
M.N. Pavlinsky\affilmark{1}, R.A. Sunyaev\affilmark{1,2} }

\affil{ 
$^1$ {\it Space Research Institute, Russian Academy of Sciences, Profsoyuznaya ul. 84/32, Moscow 117810, Russia } \\
$^2$ {\it Max Planck Institut fur Astrophysik, Karl-Schwarzschild-Str. 1, Postfach 1317, D-85741 Garching, Germany}
}

\vspace{2mm}

{\bf Abstract\/}.We investigated the optical, X-ray, and gamma-ray variability of the pulsar
\mbox{SAX J2103.5+4545}. Our timing and spectral analyses of the X-ray and gamma-ray emissions from
the source using \mbox{RXTE} and \mbox{INTEGRAL} data show that the shape of its spectrum in the energy range
3 -- 100 keV is virtually independent of its intensity and the orbital phase. Based on XMM-Newton data, we
accurately (5\arcsec ) localized the object and determined the optical counterpart in the binary. We placed upper
limits on the variability of the latter in the R band and the $H_{\alpha}$ line on time scales of the orbital and pulse
periods, respectively. 
\copyright\,{\it 2004 MAIK "Nauka/Interperiodica"}.


\vfill
 
{$^{*}$ E-mail: aal@hea.iki.rssi.ru}

\newpage
\thispagestyle{empty}
\setcounter{page}{1}

\section*{INTRODUCTION}

The X-ray transient SAX J2103.5+4545 was discovered by the BeppoSAX observatory during its outburst 
in 1997 (Hulleman {\it et al.} 1998). Almost immediately, coherent pulsations with a period of $\sim$358 s
were detected in the source, which allowed it to be classified as a transient X-ray pulsar. RXTE 
observations of the source during the next outburst in 1999 revealed another type of periodicity attributable to
the orbital motion of the compact object. Based on these observations, Baykal {\it et al}. (2000) found the
pulsar to be a member of a binary and to have an elliptical orbit with an eccentricity of $e\simeq$0.4
 and an orbital period of  $\sim$12.68\,d. Subsequently,Baykal {\it et al}. (2002) estimated the distance to the
source,$\sim$3.2 kpc, and its X-ray luminosity, $L_x\sim6\times 10^{34}-10^{36}$ erg s$^{-1}$ .

    An analysis of the light curves for the object showed that the intensity of the pulsar is highly
variable within one orbital cycle and peaks near the periastron (Baykal {\it et al}. 2000). Such a behavior of
the light curves is typical of binaries with relatively high eccentricities and high-mass companions --
early-type (O -- B) stars. Hulleman {\it et al}. (1998) suggested that the star HD\,200709 could be the
optical counterpart. However, the position of this star outside the \mbox{BeppoSAX} error region and its spectral
type (B8 V) made this candidacy highly questionable. Using the localizations of \mbox{SAX J2103.5+4545} by the
\mbox{BeppoSAX} observatory and by the IBIS and JEM-X telescopes of the INTEGRAL observatory as well
as the observations of this region at the Skinakas observatory (Crete), Reig {\it et al}. (2004) pointed to
another candidate; their error region for the sources $\sim30$\arcsec.

     In this paper, we make an attempt to investigate the variability of the source on time scales from 
several hundred seconds (the spin period of the neutron star) to several days (the orbital period of the binary)
over a wide energy range, from optical (\mbox{RTT-150}) to hard X-rays (\mbox{RXTE} and \mbox{INTEGRAL}). In addition,
using \mbox{XMM-Newton} data, we were able to increase the localization accuracy for the X-ray pulsar to  5\arcsec.

\section*{OBSERVATIONS}

\subsection*{\it RTT-150}
     Optical observations of SAX J2103.5+4545 were performed in the fall of 2003 with the Russian-
Turkish 1.5-m telescope (RTT-150, TUBITAK National Observatory, Turkey, Mount Bakyrly, 2547 m,
$2^\mathrm{h}01^\mathrm{m}20^\mathrm{s}$~E, $36^\circ49'30''$~N). The observations were carried out with
 a back-illuminated Andor Technologies 2$\times$2  CCD array placed at the Cassegrain focus of the 
telescope (1 : 7.7). A median zero-exposure frame and dark current were subtracted from all images, 
and the images were divided by a flat field.
We reduced the images using the IRAF (Image Reduction and Analysis Facility) standard software
package\footnote{http://tucana.tuc.noao.edu/} and our own software.

\subsection*{\it XMM-Newton}                      
   To improve the celestial coordinates of the source, we used data from the XMM-Newton observatory. Its
main instruments are three grazing-incidence X-ray telescopes with (MOS1, MOS2, PN) CCD arrays
placed at the foci. The typical angular size of a point source on the detectors is of the order of several 
arcseconds, which allows the coordinates of the source to be determined with a high accuracy. Here, we used
data from the EPIC MOS instruments operating in the energy range 0.15 -- 12 keV.

\subsection*{\it  INTEGRAL}
   The International gamma-ray observatory \mbox{INTEGRAL} (Winkler {\it et al}. 2003) was placed into a
high-apogee orbit by the Russian Proton Launcher from the Baikonur Cosmodrome on October 17,
2002 (Eismont {\it et al}. 2003). The payload of the satellite includes the SPI gamma-ray spectrometer,
the IBIS gamma-ray telescope, the JEM-X X-ray monitor, and the OMC optical monitor (for more
details, see Winkler {\it et al}. (2003) and references therein). We used data from the IBIS telescope, its
upper ISGRI detector (Lebrun {\it et al}. 2003), and data from the JEM-X X-ray monitor (Lund {\it et al}. 2003).
Both instruments operate on the principle of a coded aperture. The field of view is $29^\circ\times29^\circ$ (total) and
$9^\circ\times9^\circ$ (the full-coding region) for IBIS and $4^\circ$ in diameter (the full-coding region) for JEM-X.

   We used the publicly accessible INTEGRAL calibration observations of the Cyg X-1 region performed
in December 2002. Preliminary results of the analysis of the INTEGRAL observations for the pulsar
\mbox{SAX J2103.5+4545} were presented by Lutovinov {\it et al}. (2003).

   The standard OSA-3.0 software package provided by the INTEGRAL Science Data Center 
(ISDC)\footnote{http://isdc.unige.ch} was used for the timing analysis of ISGRI data and
the analysis of JEM-X data. A method described by Revnivtsev {\it et al}. (2004) was used to reconstruct the
images and to construct the spectrum of the source from ISGRI data. An analysis of the observational
data for the Crab Nebula indicates that the technique used allows the spectrum of the source to be
accurately reconstructed; the systematic uncertainty is  10\% for the absolute normalization of the flux
obtained and  5\% in each energy channel when reconstructing the spectrum of the source. The latter
was added as a systematic uncertainty in the spectral analysis of the source in the XSPEC package.

 \subsection*{\it RXTE}
    For our comparative analysis of the \mbox{INTEGRAL} results, we used the simultaneous observations of the
pulsar SAX J2103.5+4545 that were performed by the RXTE observatory (Bradt {\it et al}. 1993) 
in December 2002 (Obs. ID. 70082-02-43 -- 70082-02-52) and that are publicly accessible.

    The main instruments of the RXTE observatory are the PCA and HEXTE spectrometers that jointly
cover the energy range 3 -- 250 keV. The PCA spectrometer is a system of five xenon proportional 
counters. The PCA field of view is bounded by a circular collimator with a radius of  $1^\circ$ at half maximum, the
operating energy range is 3 -- 20 keV, the effective area at energies of 6 -- 7 keV is $\sim$ 6400 $cm^2$, and the
energy resolution at these energies is  $\sim$18\%. The HEXTE spectrometer is a system of two independent
packages of four phoswich NaI(Tl)/CsI(Na) detectors rocking with a period of 16 s for the observations
 of off-source areas at a distance of 1.5$^\circ$ from the source. At each specific time, the source can be
observed only by one of the two detector packages; thus, the effective area of the HEXTE detectors is $\sim$700 $cm^2$.
 The operating energy range of the spectrometer is 15 -- 250 keV.

    The standard FTOOLS/LHEASOFT 5.3 software package was used to reduce the RXTE data. In
our spectral analysis of the PCA data in the energy
range 3 -- 20 keV, we introduced a systematic uncertainty of 1\%.

\section*{LOCALIZATION AND DETERMINATION OF THE OPTICAL COUNTERPART}

    Using INTEGRAL observations, we obtained an image of the sky region with the source in the energy
range 18 -- 60 keV (Fig. 1). The position of the source can be determined from these data with an accuracy
of $1'$ . This accuracy is too low to unambiguously determine the optical counterpart to the source. Reig
{\it et al}. (2004) made an attempt to improve the localization accuracy using the overlapping BeppoSAX
and INTEGRAL error regions.

    Here, we used archival XMM-Newton (Obs.Id 0149550401) data to improve the position of
\mbox{SAX J2103.5+4545}. During these observations, the telescope operated in fast-variability mode, in which
information is read from the CCD array only along one of the axes. Compared to the standard modes, this
mode allows the time resolution to be increased significantly, but at the same time, it has a shortcoming:
there is no direct spatial information. To determine the celestial coordinates of the source, we used data
from two detectors, MOS1 and MOS2, in which information is read along the mutually perpendicular
directions. We used the following algorithm:

\begin{enumerate}

\item We determined the detector coordinate of the
centroid of the one-dimensional photon distribution
for each of the detectors (RAWX1$_{src}$, RAWX2$_{src}$).
\item We generated a set of celestial coordinates in
the region where the source was presumably located.
Subsequently, this set was transformed into two sets
of detector coordinates (one for MOS1 and the other
for MOS2) using the standard esky2det code from
the Science Analysis System (SAS) of the XMM
observatory.
\item From the set of celestial coordinates, we chose
those for which the detector coordinates (RAWX1,
RAWX2) corresponded to those of the source
(RAWX1$_{src}$, RAWX2$_{src}$).

\end{enumerate}
   As a result, we obtained the following coordinates
of the source: $\alpha = 21^h03^m 36^s$, $\delta = 45^d 45^m 07^s$.
The localization accuracy is determined by the response function of the telescope, the astrometric
referencing accuracy, and the peculiarities of our coordinate determination procedure. We obtained 
 $\sigma_{RADEC}=5$\arcsec as an estimate of the total error. 

The RTT-150 map of the sky region around the
pulsar \mbox{SAX J2103.5+4545} is shown in Fig. 2. The circles indicate the error regions determined from
BeppoSAX and XMM-Newton data. It clearly follows from this figure that the optical counterpart to
\mbox{SAX J2103.5+4545} is determined with a high degree of confidence from the results of our analysis. It is
the star shown in Fig. 2. This result agrees with that obtained by Reig {\it et al}. (2004). The RTT-150
measurements yield the magnitudes R = 13.60 and V = 14.20 for this star. The derived color is consistent 
with the emission from an O-B star at an interstellar reddening of A$_V$ = 3.12.

\section*{TIMING ANALYSIS}

\subsection*{Variability on Time Scales of the Orbital Period}
   Lutovinov {\it et al}. (2003) showed that the intensity of the source in the hard energy ranges 15 -- 40 and
40 -- 100 keV is highly variable and depends on the orbital phase. Figure 3a presents the light curve of
the pulsar constructed from ISGRI data in the energy range 18 -- 60 keV. Each point was obtained by averaging
 over five individual pointings and has an exposure time of  $\sim15$ ks. To depict the dependence of the
source's intensity more conveniently, orbital phases are plotted along the horizontal axis together with
time (the parameters of the binary were taken from Baykal {\it et al}. (2000)). The flux from the binary is at
a maximum ($\sim$ 40 -- 50 mCrab) near orbital phases of 0.55 -- 0.7 and decreases to $\sim$10 -- 20 mCrab at phases
of 0.1 -- 0.2. As was noted by Lutovinov {\it et al}. (2003), the observed hard X-ray light curve follows almost
closely the light curve in the standard X-ray energy range (Baykal {\it et al}. 2000).

   The variability of the source was also studied at optical wavelengths on time scales of the order of
the orbital period, $\sim$12.6 days. For this purpose, the source's field was imaged in the R band (the band was
chosen arbitrarily) in the second half of October and November 2003 on each night, where possible. The
derived light curve is shown in Fig. 3b. We found no variability of the source related to its orbital motion in
the binary; the upper limit on its amplitude is  1\%.

\subsection*{Variability on Time Scales of the Pulsation Period}
   The periods of X-ray pulsars are known to be variable and subject to both long-term changes and
small-scale fluctuations (see, e.g., Nagase 1989; Lutovinov {\it et al}. 1994; Bildsten {\it et al}. 1997). While 
monitoring SAX J2103.5+4545 during its 1999 outburst, Baykal {\it et al}. (2002) found a significant spin-up of
the neutron star, with the observed spin-up rate being proportional to the flux from the source. Figure 4
shows the changes in the pulsar's period with time throughout the history of its observations by different
observatories.

   An epoch-folding technique was used to determine the pulsar's period during the INTEGRAL observations. 
The source's light curve in the energy range 20 -- 100 keV with a time step of 40 s was constructed from ISGRI data 
using standard software and corrected for the orbital motion of the neutron star
in the binary using known orbital parameters (Baykal et al. 2000). The pulsation period calculated in this
way was 355.10$\pm$0.04 s. Figure 5 shows the $\chi^2$ periodogram for the source's light curve obtained by
searching for flux pulsations from it. The error in the period was determined by the Monte Carlo method
from an analysis of the simulated light curves.  

  Figure 6a shows the phase light curve for the
pulsar SAX J2103.5+4545 constructed from \mbox{INTEGRAL} data in the energy range 20 -- 100 keV. It has
a single-peaked shape extended over the entire phase cycle of the pulsar's period. The intensity rapidly rises
and smoothly decays. Baykal {\it et al}. (2000) and Inam {\it et al}. (2004) provided the source's pulse profiles for
different soft X-ray energy ranges. Our hard X-ray pulse profile slightly differs from the previous soft 
X-ray profiles, whose peaks occupy only half of the cycle, and the rise in intensity is smoother than its decay.
Based on the INTEGRAL data reduction results, we failed to estimate the pulse fraction in the hard
energy range, because the standard software used to construct the light curves of sources does not 
 properly estimate the contribution from the background radiation on the detector. Therefore, to estimate this
contribution, we used the HEXTE/RXTE data obtained over the same time interval as the INTEGRAL 
data. According to these data, the pulse fraction is  $\sim20\pm$5\%. In the soft energy range 0.9 -- 11 keV, the
pulse fraction is $50.9\pm0.3$\% (Inam {\it et al}. 2004).

    The variability of the source on time scales of its X-ray pulsations was also studied at optical 
 wavelengths, in an $H_{\alpha}$ filter. The observations of such variability in other high-mass binaries with pulsars
were reported previously (e.g., in the object X Persei; Mazeh {\it et al}. 1982). The $H_{\alpha}$ observations of the
source were performed on November 18, 2003 (52961 MJD), with the RTT-150 telescope. The observations were 
carried out for 1.1 h with a time resolution of $\sim$15 s. 

Since the source has exhibited a nonuniform spin-up 
 of the neutron star throughout the history of its observations, to calculate the expected period at the
epoch of our optical observations, we have obtained a conservative estimate of the mean spin-up rate
as follows. For all of the possible pairs of points in Fig. 4, we calculated the spin-up rate between
them and then determined its mean value,  $\dot P/P\sim3.2\times 10^{-3}$ yr$^{-1}$. In this procedure, we excluded the
period measured by the INTEGRAL observatory in the series of observations in May -- June 2003 
(Sidoli {\it et al}. 2003). 
This decision was justified in part by the large uncertainty in the measure period at this epoch
and by the fact that, if the presumed spin-up rate of the pulsar was estimated from our measurements and
from the measurements by Inam {\it et al}. (2004) and Sidoli {\it et al}. (2004), then the value obtained would
be several times higher than the maximum spin-up rate observed by Baykal {\it et al}. (2002) during the 1999
outburst. It is important to note that the 3 -- 20 keV flux from the source in December 2003 -- April 2004
was comparable to its flux in 1999 -- 2000. The ultimate answer to the question concerning the behavior
of the spin-up rate of the neutron star in the binary \mbox{SAX J2103.5+4545} may be given after analyzing the
large set of RXTE observations of this object in 2003 that is not yet publicly accessible.

    The presumed pulsation period at the epoch of our optical observations estimated by the method described above
 is 354.02 s. The $H_{\alpha}$  light curve of the optical counterpart folded with this period is shown in
Fig. 6b. We found no variability of the source in then $H_{\alpha}$  band; the upper limit is  1\%. The zero phase of 
the X-ray pulse profile and the presumed zero phase at the epoch of our optical observations calculated
using the procedure described above were brought   into coincidence.

\section*{SPECTRAL ANALYSIS}

As we noted above, the intensity of the source varies greatly with the orbital motion of the neutron 
star in the binary. Therefore, we performed a spectral analysis of the pulsar emission for various
orbital phases of the binary with the goal of finding the possible dependence of the source's spectrum on
its intensity and its position in the orbit. Such an analysis was performed by Baykal {\it et al}. (2002) using
RXTE data for the standard X-ray energy range 3 -- 20 keV. In contrast, we investigated the behavior of
the source over a wide energy range up to  100 keV using INTEGRAL and RXTE data.

We used JEM-X and ISGRI data to construct the spectrum of the source from its INTEGRAL observations in the energy 
ranges 6 -- 20 and 20 -- 100 keV, respectively. To test the validity of the normalization
of the JEM-X spectra, we analyzed a series of spectra for the Crab Nebula. Where possible, we took observations 
when this object was within the same areas of the JEM-X field of view as our source. We found
that the shape of the spectrum for the Crab Nebula is reconstructed satisfactorily, while the normalization
proves to be underestimated by a factor of  $\sim1.6$. This factor was used for the correction of the normalization 
of the JEM-X spectrum for the source. Our analysis of the source's spectrum at various orbital
phases revealed no appreciable deviations of its shape over a wide energy range either. This allowed us to
subsequently study the source's average spectrum shown in Fig. 7. For comparison, this figure also
shows the pulsar's spectrum constructed from the RXTE observations performed over the same period
as the INTEGRAL observations. We used PCA and HEXTE detector data for the energy ranges 4 -- 20
and 20 -- 70 keV, respectively. It should be noted that the total time of RXTE observations of the source is
 shorter than the time of its INTEGRAL observations. 

The spectrum of the X-ray pulsar
\mbox{SAX J2103.5+4545} is typical of this class of objects and can be described by a simple power law with
an exponential high-energy cutoff. This model has long and widely been used to fit the spectra of X-ray 
pulsars (White {\it et al}. 1983). Based on \mbox{XMM-Newton} data, Inam {\it et al}. (2004) measured the
neutral hydrogen column density $N_H$. Depending on the model used to describe the pulsar's spectrum, this 
parameter varies over the range $N_H=(0.6-0.9)\times10^{22}$ atoms cm$^{-2}$, in agreement with
our optical measurements (see above). Since there are no INTEGRAL data at energies below 6 keV in
our case and since the RXTE sensitivity is lower (than the  XMM-Newton sensitivity) at soft  energies, we
fixed $N_H$ at $0.9\times10^{22}$ atoms cm$^{-2}$ in the subsequent analysis.

   The table gives the best-fit parameters for the model fit to the source's spectrum described above
for the INTEGRAL and RXTE data. The derived parameters for the two data sets are in good agreement
between themselves and with the values from Baykal {\it et al}. (2002) and Inam {\it et al}. (2004), except for the
e-folding energy that proved to be slightly lower. This may be because we greatly extended the energy range
under study. Note also that the RXTE spectrum of the source exhibits a neutral iron line at an energy of
6.4 keV (see the table).

\section*{CONCLUSIONS}

    The transient X-ray pulsar \mbox{SAX J2103.5+4545} is a member of a high-mass Be binary with a moderate
eccentricity and the shortest orbital period known to date among all such binaries. Identifying the optical
counterpart of the binary plays one of the most important roles in understanding the nature of binaries with
compact objects and the processes that take place in them.

    We have been able to localize the pulsar with an accuracy of  5\arcsec and to firmly establish its nature:
an emission-line O-B star. We performed optical observations of this object with the RTT-150 telescope
with the goal of finding its possible variability. We found no variability of the optical counterpart on time
scales of the orbital period and the spin period of the pulsar; the upper limit on its amplitude is about 1\% in
both cases.

  The X-ray pulsar has exhibited a spin-up almost throughout the history of its observations,
with the spin-up rate depending on the luminosity of the source (Baykal {\it et al}. 2002). During the \mbox{INTEGRAL} 
observations, the pulsation period was 355.10$\pm$0.04 s. The pulse fraction decreases with
increasing energy and is  20\% in the energy range 20 -- 100 keV.

    Our analysis of the INTEGRAL and RXTE observational data for the pulsar shows that the source
is detected at a statistically significant level up to energies of $\sim$100 keV. As in the standard X-ray energy
range, the intensity of the source in the hard energy range also depends on the orbital phase of the binary
and peaks near the periastron. At the same time, the shape of the source's spectrum remains virtually unchanged over 
 a wide energy range (3 -- 100 keV) and can be described by the standard (for X-ray pulsars)
model: a simple power law with a high-energy cutoff and low-energy absorption.

    We detected no features in the 3 -- 100 keV spectrum of the source that could be interpreted as cyclotron lines.
 Thus, we can impose lower and upper limits on the magnetic field strength of the source: B > 8.6$\times10^{12}$ G 
and B < 2.6$\times10^{11}$ G, respectively. We know X-ray pulsars with a weak magnetic field ($B\sim10^{11}$ G), 
for example, SMC X-1 (see, e.g., Lutovinov {\it et al}. (2004) and references therein) and
GRO J1744-28 (Rappaport and Joss 1997). The detection of type II X-ray bursts from them provides
evidence that the magnetic field in these sources is weak. No such bursts are observed from the pulsar
SAX J2103.5+4545. Baykal {\it et al}. (2002) estimated the magnetic field from the relationship between the
rate of change in the pulsar's period and its parameters (Ghosh and Lamb 1979): $B\sim12\times10^{12}$ G.
Thus, most of the arguments suggest that the neutron star in the binary under consideration has a
strong magnetic field.

\section*{ACKNOWLEDGMENTS}
   We thank E.M. Churazov, who developed the algorithms for IBIS data analysis and provided the
software. We also thank S.V. Molkov and S.S. Tsygankov for help in reducing the JEM-X and RXTE
data. This work was supported by the Ministry of Industry and Science (Presidential grant no. NSh2083.2003.2 
and project no. 40.022.1.1.1102) and the Russian Foundation for Basic Research (project
no. 04-02-17276). We are grateful to the INTEGRAL Science Data Center (Versoix, Switzerland)
and the Russian INTEGRAL Science Data Center (Moscow, Russia). The results of this work are
based on observations of the INTEGRAL observatory, an ESA project with the participation of Denmark, 
France, Germany, Italy, Switzerland, Spain, Czechia, Poland, Russia, and USA. We also used
data retrieved from the High-Energy Astrophysics Archive, The Goddard Space Flight Center. A.A.L.,
M.G.R., R.A.B., and M.N.P. thank the International Space Science Institute (ISSI, Bern) for support.

\newpage

\section*{REFERENCES}

\begin{enumerate}

\item A. Baykal, M. Stark, and J. Swank,Astrophys.J. Lett. 544, L129 (2000).

\item A. Baykal, M. Stark, and J. Swank, Astrophys. J. 569, 903 (2002).

\item L. Bildsten, D. Chakrabarty, J. Chiu, et al., Astron. Astrophys., Suppl. Ser. 113, 367 (1997).

\item H. V. Bradt, R. E. Rothschild, and J. H. Swank,
    Astron. Astrophys., Suppl. Ser, 97, 355 (1993).

\item N. Eismont, A. Ditrikh, G. Janin, et al., Astron. Astrophys. 411, L37 (2003).

\item P. Ghosh and F. Lamb, Astrophys. J. 234, 296 (1979).

\item F. Hulleman, J. in`t Zand, and J. Heise, Astron. Astrophys. 337, L25 (1998).

\item S. Inam, A. Baykal, J. Swank, et al., Astrophys. J.(2004, in press); astro-ph/0402221.

\item F. Lebrun, J. P. Leray, P. Lavocat, et al., Astron. Astrophys. 411, L141 (2003).

\item N. Lund, S. Brandt, C. Budtz-Joergesen, et al., Astron. Astrophys. 411, L231 (2003).

\item A. A. Lutovinov, S. V. Molkov, and M. G. Revnivtsev,
    Pis'ma Astron. Zh. 29, 803 (2003) [Astron. Lett. 29,713 (2003)].

\item A. A. Lutovinov, S. S. Tsygankov, S. A. Grebenev, et
    al., Pis'ma Astron. Zh. 30, 58 (2004) [Astron. Lett.30, 50 (2004)].

\item T. Mazeh, R. Treffers, and S. Vogt, Astrophys. J. 256, L13 (1982).

\item F. Nagase, Publ. Astron. Soc. Jpn. 41, 1 (1989).

\item A. Rappaport and P. Joss, Astrophys. J. 486, 435 (1997).

\item R. Reig, I. Negueruela, J. Fabregat, et al., Astron. Astrophys. 421, 673 (2004).

\item M. G. Revnivtsev, R. A. Sunyaev, D. A. Varshalovich,
    et al., Pis'ma Astron. Zh. 30, 430 (2004) [Astron.Lett. 30, 382 (2004)].

\item L. Sidoli, S. Mereghetti, S. Larsson, et al., Proceeding of the 5th INTEGRAL Workshop (2004, in
    press); astro-ph/0404018.

\item N. White, J. Swank, and S. Holt, Astrophys. J. 270,
    771 (1983).

\item C. Winkler, T. J.-L. Courvoisier, D. Di Cocco, et al.,Astron. Astrophys. 411, L1 (2003).

\end{enumerate}
             Translated by V. Astakhov

\newpage

\begin{center}
{\bf Table}{Best-fit parameters for the spectrum of the pulsar SAX J2103.5+4545}\\

\begin{tabular}{l|l}
\hline 
\hline

Parameters &Values \\
&\\
\hline

\multicolumn{2}{c}{Derived from INTEGRAL data}\\

\hline
 &\\
$N_H$,$10^{22}$cm$^{-2}$ &0.9 (fixed) \\
 &\\
Photon index $\Gamma$  &1.04$\pm$0.15 \\
&\\
Cutoff energy $E_{cut}$,keV &8.5$\pm$2.4 \\
&\\
{\it e}-folding energy $E_{fold}$,keV &21.37$\pm$2.75 \\
&\\
$\chi^2$(degree of freedom) & 1.21\\
\hline
\hline
 
\multicolumn{2}{c}{Derived from RXTE data}\\

\hline
 &\\
$N_H$,$10^{22}$ &0.9 (fixed) \\
 &\\
Photon index $\Gamma$ &0.979$\pm$0.066 \\
&\\
Cutoff energy $E_{cut}$,keV  &6.97$\pm$1.26 \\
&\\
{\it e}-folding energy $E_{fold}$,keV  &22.98$\pm$1.73\\
&\\
Fe line center, keV &6.34$\pm$0.32\\
&\\
Fe line width, keV &0.81$\pm$0.21\\
&\\
Fe line intensity, photons cm$^{-2}$ s$^{-1}$ &(1.2$\pm$0.4)$\times10^{-3}$\\
&\\
$\chi^2$(degree of freedom) & 0.99\\
\hline

\end{tabular}

\end{center}

\newpage

\begin{figure*}[t]
  \includegraphics[width=17cm]{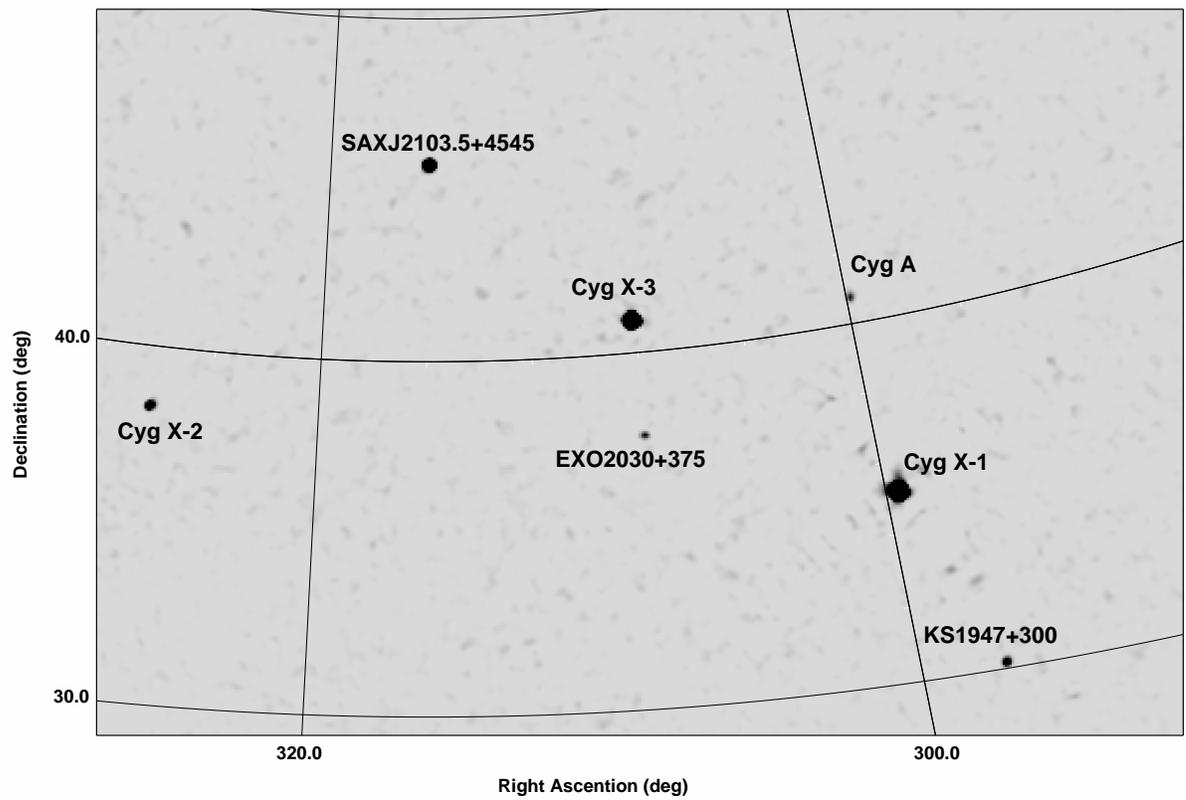}

\vfill
\renewcommand{\figurename}{Fig.} 

\caption{The region of ISGRI observations of SAX J2103.5+4545 in the energy range 18 -- 60 keV.}
\end{figure*}
\newpage

\begin{figure*}[t]
  \includegraphics[width=17cm]{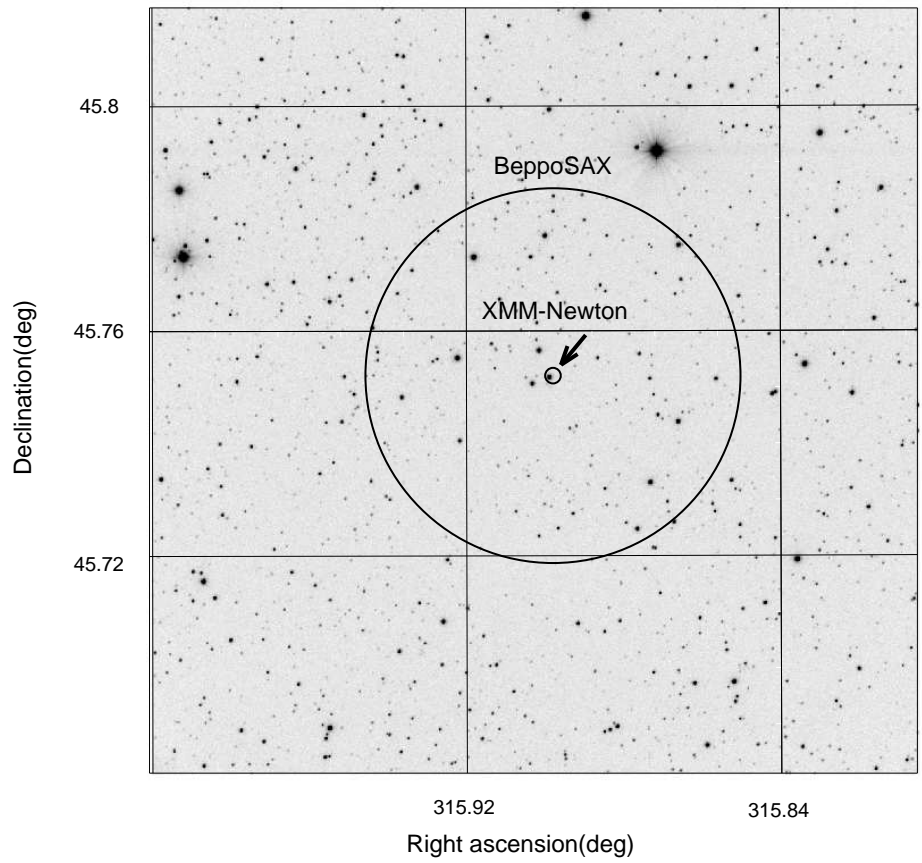}
\vfill
\renewcommand{\figurename}{Fig.}
\caption{The region of ground-based RTT-150 R-band observations of SAX J2103.5+4545. The arrow indicates the optical counterpart to the pulsar, and its error region determined here from XMM-Newton data is highlighted.}
\end{figure*}
\newpage

\begin{figure*}[t]
  \includegraphics[width=17cm,clip]{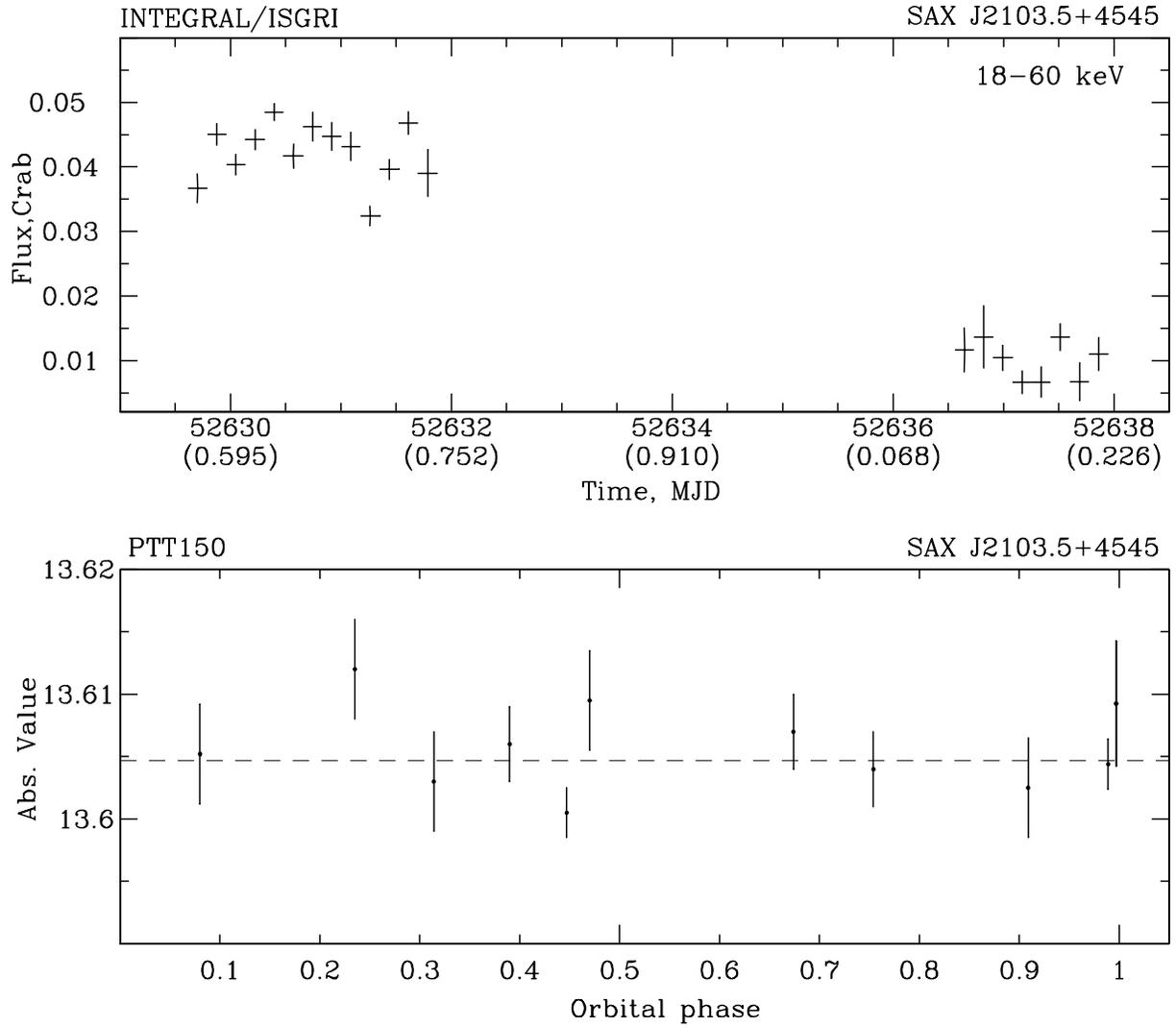}

\vfill
\renewcommand{\figurename}{Fig.}
\caption{Light curve of the pulsar SAX J2103.5+4545 on time scales of the orbital period: constructed from ISGRI data in the energy range 18 -- 60 keV (upper panel) and from ground-based RTT-150 R-band observations in October -- November 2003 (lower panel). For convenience, optical phases are plotted along the horizontal axis. The errors correspond to one standard deviation. The dashed line indicates the mean R magnitude of the star.
}
\end{figure*}

\newpage

\begin{figure*}[t]
  \includegraphics[width=17cm,clip]{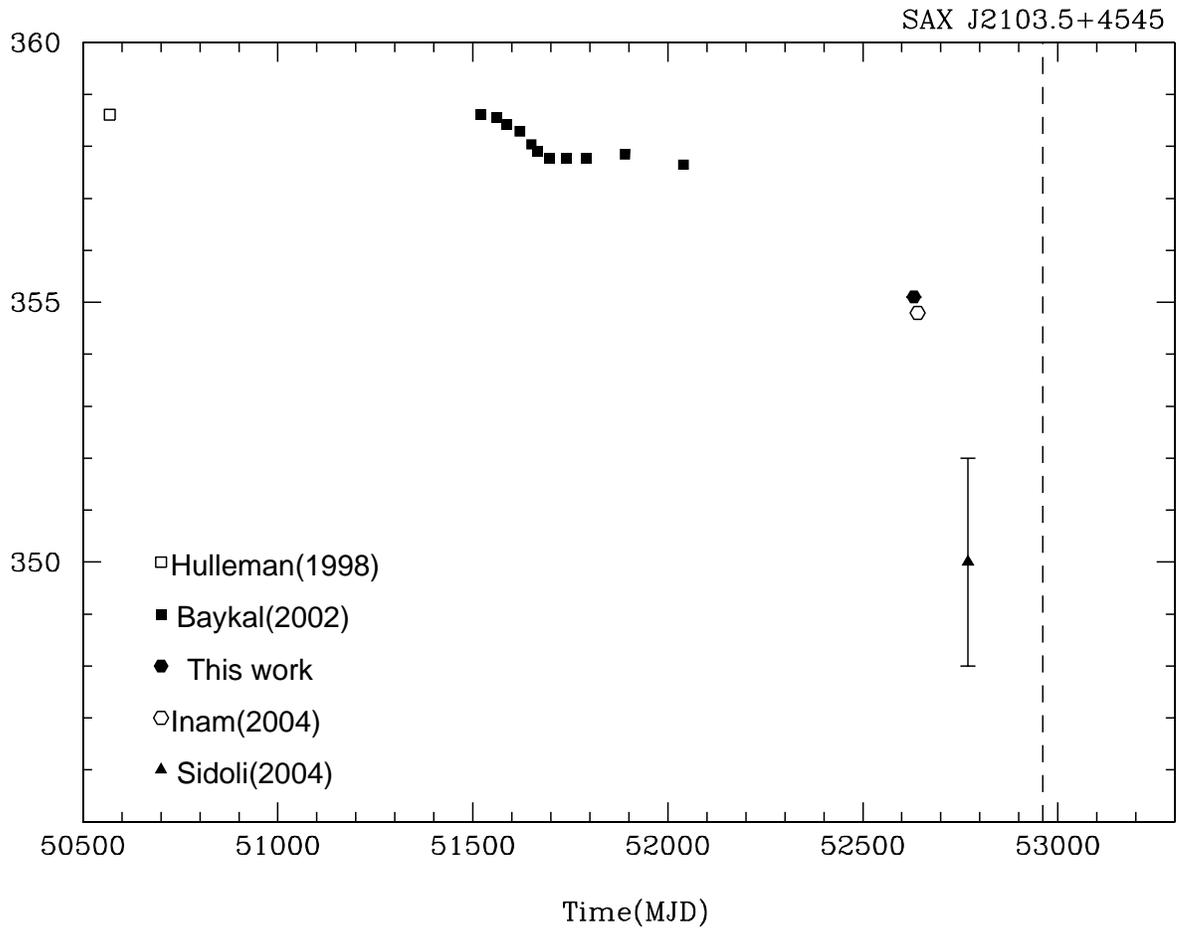}

\vfill
\renewcommand{\figurename}{Fig.}
\caption{Changes in the pulsation period of SAX J2103.5+4545 throughout the history of its observations by different observatories. The dashed line marks the epoch of optical observations.
}
\end{figure*}

\newpage

\begin{figure*}[t]
  \includegraphics[width=15cm]{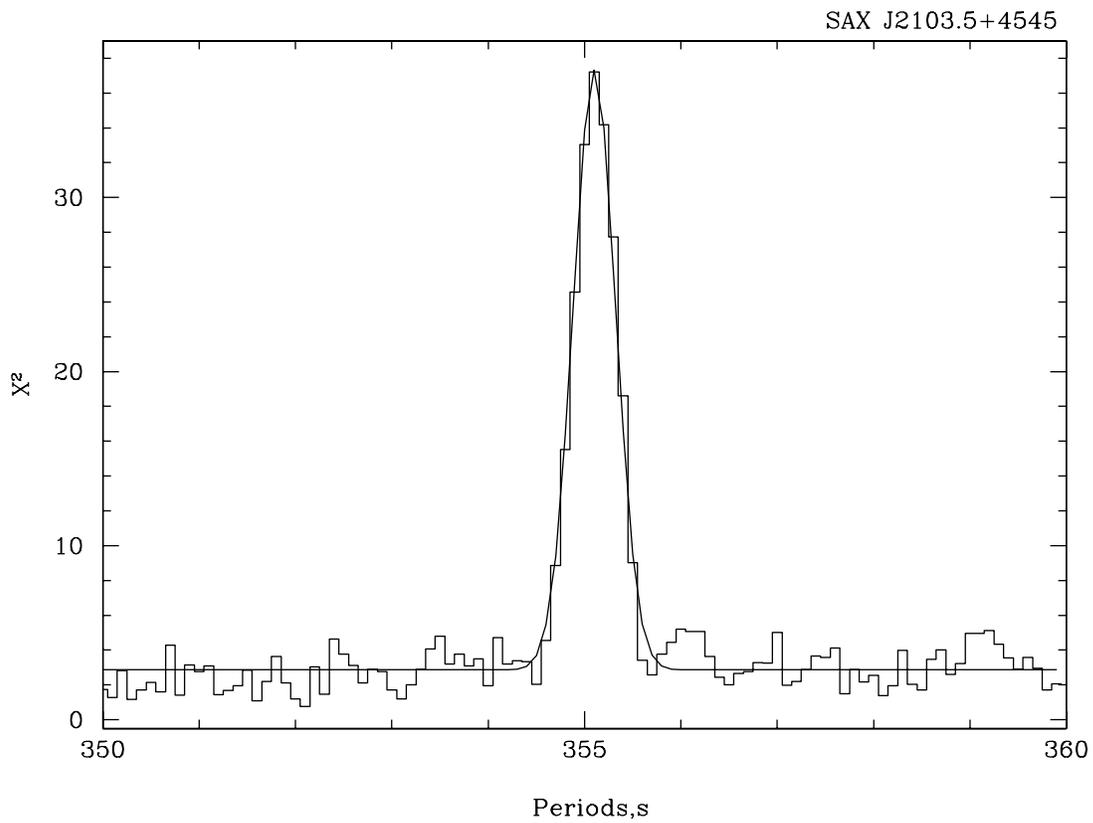}

\vfill
\renewcommand{\figurename}{Fig.}
\caption{$\chi^2$ periodogram obtained by searching for flux pulsations from SAX J2103.5+4545 by the epoch-folding technique. The solid line indicates the Gaussian best fit.
}
\end{figure*}

\newpage

\begin{figure*}[t]
  \includegraphics[width=17cm]{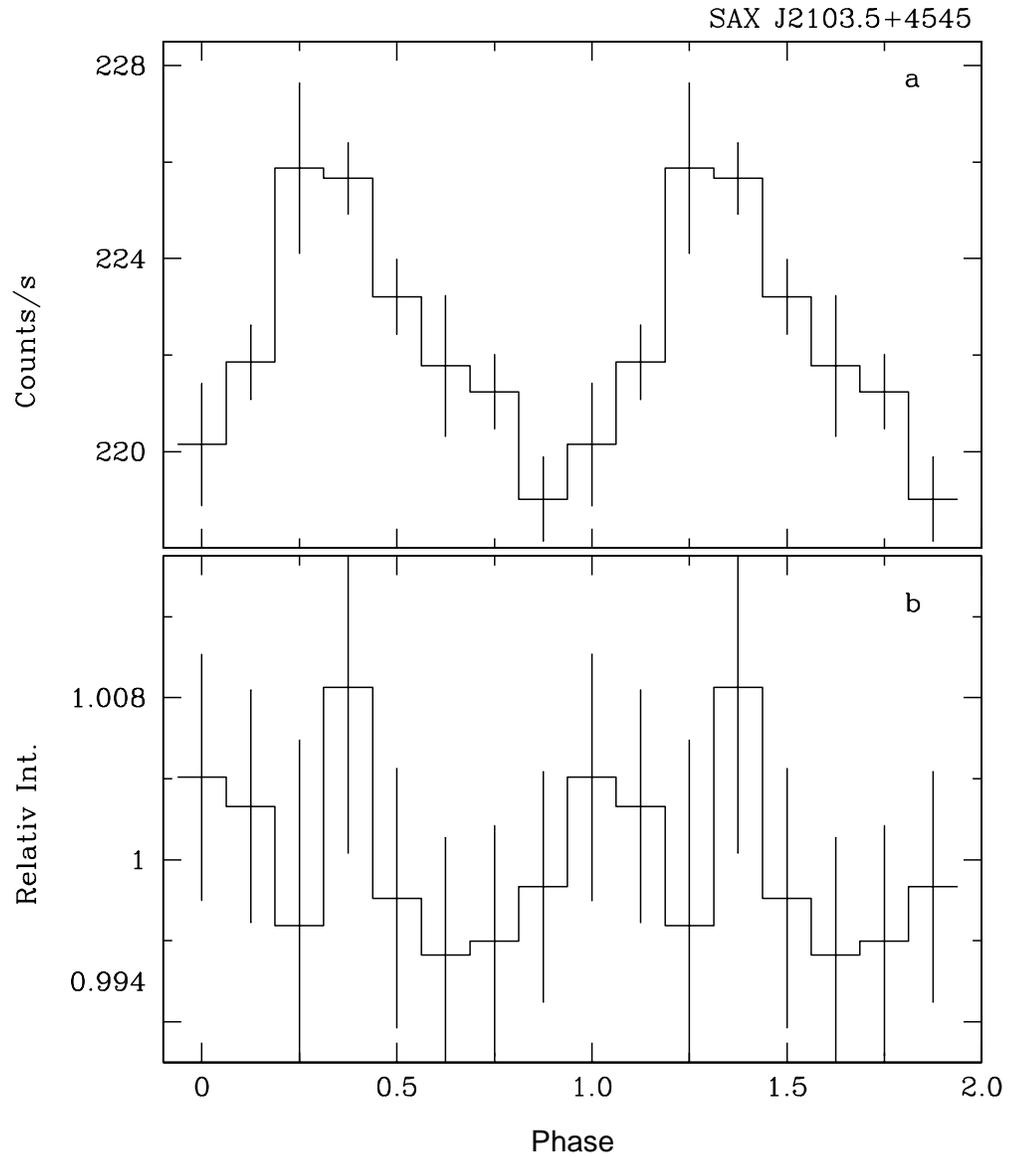}

  \vfill
\renewcommand{\figurename}{Fig.}
\caption{Pulse profiles for the pulsar SAX J2103.5+4545 in the energy range 20 -- 100 keV as constructed from
IBIS/ISGRI data (the background was not subtracted) (a) and in an $H_\alpha$ filter as constructed from the RTT-150
observations on November 18, 2003 (b).}
\end{figure*}

\newpage

\begin{figure*}
  \includegraphics[width=17cm,clip]{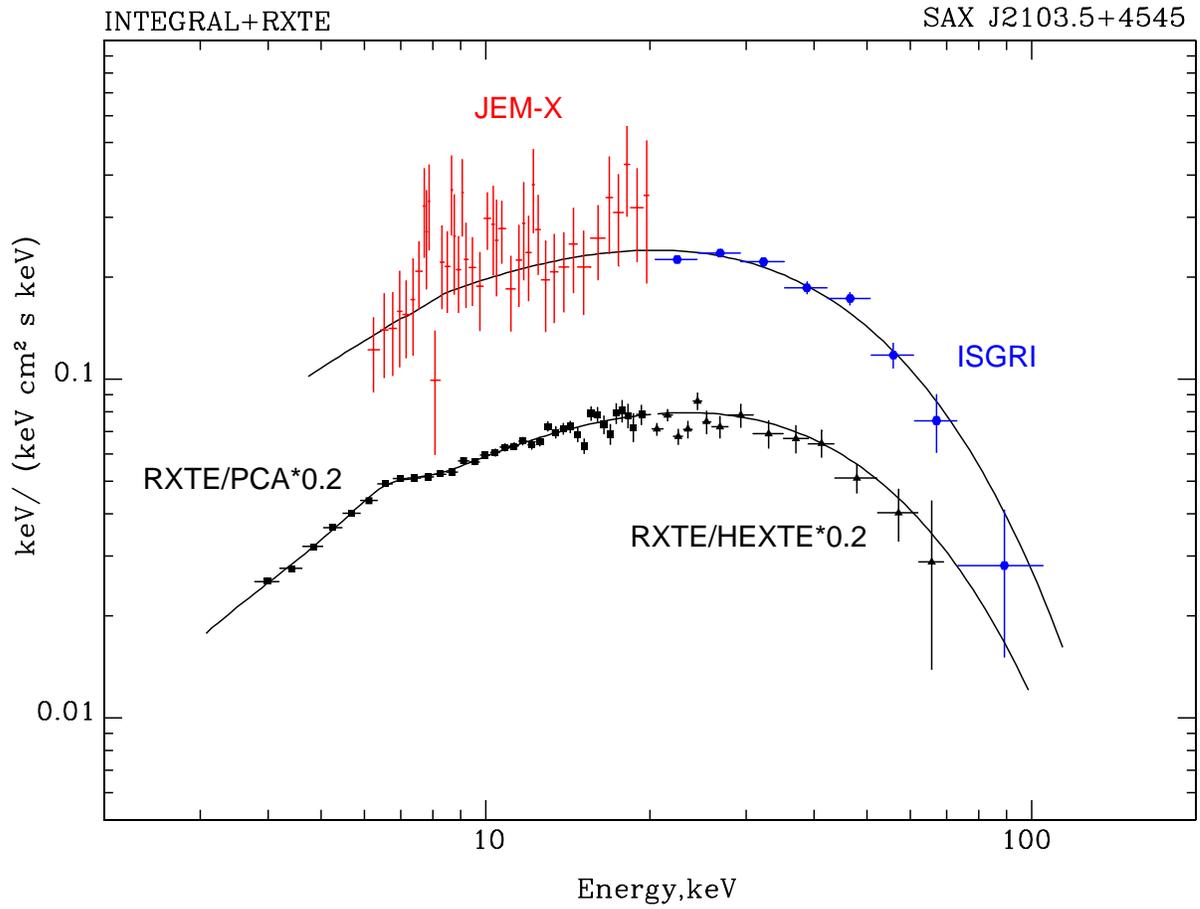}

\renewcommand{\figurename}{Fig.}
\caption{Energy spectrum of the pulsar SAX J2103.5+4545 as constructed from INTEGRAL and RXTE data. The normalization of the RXTE spectrum was multiplied by 0.2. The lines indicate the model fits to the spectra with the best-fit parameters (see the table).
}
\end{figure*}

\end{document}